# Universal properties of cuprate superconductors

T. Schneider

*Physik-Institut der Universiät Zürich, Winterthurerstrasse 190,*

*CH-8057 Zürich, Switzerland*

**Abstract**

To provide an understanding of the universal properties emerging from the empirical correlations and phase diagrams of cuprate superconductors , we invoke the scaling theory of finite temperature and quantum critical phenomena. The universal features are traced back to the existence of quantum critical lines, representing the end lines of the finite temperature transition surface. At the respective quantum critical lines two dimensional superconductor to insulator (2D-QSI) transitions and three dimensional superconductor to normal state (3D-QSN) transitions occur. The flow to this quantum critical points is tuned by doping, substitution and anisotropy. It is shown that the empirical correlations, like the dependence of $T_c$ on dopant and substitution concentration, the dependence of $T_c$ on zero temperature in-plane penetration depth, etc., reflect universal properties associated with the flow to these quantum critical points and of the crossover from one to the other. A detailed account of the flow to 2D-QSI and 3D-QSN criticality is a challenge for microscopic theories attempting to solve the puzzle of superconductivity in these materials.

The evidence for and the implications of critical fluctuations near the superconducting phase transition have been revisited quite often since the discovery of materials which exhibit high temperature superconductivity. Although it has been difficult to confront experiment with the zoo of microscopic theories and models, there have been notable successes in understanding the phenomenology of zero and finite temperature phase transitions into the superconducting phase. Here, progress has been made employing the scaling theory of finite temperature and quantum critical phenomena [1,2]. Cuprate superconductors exhibit very rich phase diagrams, and a microscopic theory should not only explain the occurrence of superconductivity, but also explain the phase diagrams and the empirical correlation between various properties, which appear to be consistent with the universal properties characterizing fluctuation dominated phase transitions. In this review I concentrate on the analyses of empirical correlations and phase diagrams in terms of the theory of critical phenomena.

In superconductivity is derived from the insulating and antiferromagnetic parent compounds by partial substitution of ions or by adding or removing oxygen. For instance $La_2CuO_4$ can be doped either by alkaline earth ions or oxygen to exhibit superconductivity. The empirical phase diagram of $La_{2-x}Sr_xCuO_4$ [3-11], depicted in Fig.1, shows that after passing the so called underdoped limit, $x_u \approx 0.047$, $T_c$ reaches its maximum value $T_{cm}$ at $x_m \approx 0.16$. With further increase of $x$, $T_c$ decreases and finally vanishes in the overdoped limit $x_o \approx 0.273$. This phase transition line is thought to be a generic property of cuprate superconductors [12] and is well described by the empirical relation

$$T_c(x) = \frac{4T_{cm}}{x_o + x_u}(x - x_u)(x_o - x), \quad x_m = \frac{x_u + x_o}{2} \tag{1}$$

proposed by Presland [13]. Approaching the endpoints along the axis $x$, $La_{2-x}Sr_xCuO_4$ undergoes at zero temperature doping tuned quantum phase transitions. As their nature is concerned, resistivity measurements [5,14] reveal a quantum superconductor to insulator (QSI) transition in the underdoped limit and in the overdoped limit a quantum superconductor to normal state (QSN) transition [2,14].

Another essential experimental fact is the doping dependence of the anisotropy. In tetragonal cuprates it is defined as the ratio $\gamma = \xi_{ab}/\xi_c$ of the correlation lengths parallel $\xi_{ab}$ and perpendicular to $CuO_2$ layers (ab-planes). In the superconducting state it can also be expressed as the ratio $\gamma = \lambda_c/\lambda_{ab}$ of the London penetration depths due to supercurrents flowing perpendicular $\lambda_c$ and parallel $\lambda_{ab}$ to the ab-planes. Approaching a non superconductor to superconductor transition $\xi$ diverges, while in a superconductor to non superconductor transition $\lambda$ tends to infinity, while $\gamma$ remains finite as long as the system exhibits anisotropic but genuine 3D behavior. There are two limiting cases: $\gamma = 1$ characterizes isotropic 3D- and $\gamma = \infty$ 2D-critical behavior. In the Ginzburg-Landau description of layered superconductors the anisotropy is related to the interlayer coupling. The weaker this coupling is, the larger $\gamma$ is. The limit $\gamma = \infty$ is attained when the bulk superconductor corresponds to a stack of independent slabs of thickness $d_s$. With respect to experimental work, a considerable amount of data is available on the chemical composition dependence of $\gamma$. At $T_c$ it can be inferred from resistivity ($\gamma = \sqrt{\rho_c / \rho_{ab}}$ and magnetic torque measurements, while in the superconducting state it follows from magnetic torque and penetration depth ($\gamma = \lambda_c / \lambda_{ab}$) data. In Fig.1 we included the doping dependence of $\gamma_T$ evaluated at $T_c$ ($\gamma_{T_c}$) and $T = 0$ ($\gamma_{T=0}$). As the dopant concentration is reduced, $\gamma_{T_c}$ and $\gamma_{T=0}$ increase systematically and tend to diverge in the underdoped limit. Thus the temperature range where superconductivity occurs shrinks in the underdoped regime with increasing anisotropy. This competition between anisotropy and superconductivity raises serious doubts whether 2D mechanisms and models, corresponding to the limit $\gamma = \infty$, can explain the essential observations of superconductivity in the cuprates. From Fig.1 it is also seen that $\gamma_T$ is well described by

$$\gamma_T = \frac{\gamma_{T,0}}{x - x_u}. \tag{2}$$

Having also other cuprate families in mind, it is convenient to express the dopant concentration $x$ in terms of $T_c$. From Eqs.(1) and (2) we obtain

$$\frac{T_c}{T_{cm}} = \frac{\gamma_{Tm}}{\gamma_T}\left(2 - \frac{\gamma_{Tm}}{\gamma_T}\right), \quad \gamma_{Tm} = \frac{\gamma_{T,0}}{x_m - x_u}. \tag{3}$$

Provided that this empirical correlation is not merely an artifact of La2-xSrxCuO4 it gives a universal perspective on the interplay of anisotropy and superconductivity, among the families of cuprates, characterized by $T_{cm}$ and $\gamma_{Tm}$. For this reason it is essential to explore its generic validity. In practice, however, there are only a few additional compounds, including HgBa2CuO4+y and Bi2Sr2CuO6+y, for which the dopant concentration can be varied continuously throughout the entire doping range. It is well established, however, that the substitution of magnetic and nonmagnetic impurities, depress $T_c$ of cuprate superconductors very effectively [15,16]. To compare the doping and substitution driven variations of the anisotropy, we depicted in Fig.2 the plot $T_c/T_{cm}$ versus $\gamma_{Tm}/\gamma_T$ for a variety of cuprate families. The collapse of the data on the parabola, which is the empirical relation (3), reveals that this scaling form appears to be universal. Thus, for a family of cuprate superconductors it gives a universal perspective on the interplay between anisotropy and superconductivity, as well as on the flow from 3D-QSN to 2D-QSN criticality with decreasing anisotropy. To shed some light on this flow it is helpful to look at the $T_c(x,y)$ phase diagram of $La_{2-x}Sr_xCu_{1-y}Zn_yO_4$ depicted in Fig.3. For any doping level where superconductivity occurs ($0.047 \leq x \leq 0.273$) the transition temperature is reduced upon Zn substitution and vanishes along the line $y_c(x)$ of quantum phase (QP) transitions. The temperature dependence of the resistivity shows that for $x < 0.16$ Zn substitution driven 2D-QSI [5] and for $x > 0.16$ 3D-QSN transitions occur [14]. The interpretation of the empirical correlation depicted in Fig.2 is now quite clear: the solid branch represents the flow to 2D-QSI- and the dashed one to 3D-QSN criticality. In the phase diagram displayed in Fig.2, these flows occur on the phase transition surface $T_c(x,y)$, where $x$ is the hole and $y$ measures the substituent concentration, e.g. for Cu, isotope substitution (using for Cu pure $^{63}$Cu or $^{65}$Cu, or for O pure $^{16}$O or $^{18}$O), or the strength of disorder. Quite generally, isotope substitution and disorder lower $T_c$ [2], while it increases initially, when hydrostatic pressure is applied [24]. Closed to optimum doping the critical behavior of the finite temperature transitions at $T_c(x,y)$ into the superconducting phase appears to be consistent with 3D-XY-critical point universality [1,2]. The universal properties of this universality class are characterized by a set of critical exponents, describing the asymptotic behavior of the correlation length $\xi_i$, magnetic penetration depth $\lambda_i$, specific heat $A^{\pm}$, in terms of

$$\xi_i^{\pm} = \xi_{i0}^{\pm} |t|^{-\nu}, \quad \lambda_i = \lambda_{i0} |t|^{-\nu/2}, \quad c = \frac{A^{\pm}}{\alpha} |t|^{-\alpha}, \quad t = T/T_c - 1, \ldots\ldots (4)$$

where $3\nu = 2-\alpha$ [1,2]. As usual, in the above expression $\pm$ refer to $t>0$ and $t<0$, respectively. The critical amplitudes $\xi_{i0}^{\pm}$, $\lambda_{i0}$, $A^{\pm}$, etc., are no universal, but there are universal critical amplitude relations, including [1,2],

$$\left(k_B T_c\right)^3 = \left(\frac{\Phi_0^2}{16\pi^3}\right)^3 \frac{\left(R^-\right)^3}{A^- \lambda_{a,0}^2 \lambda_{b,0}^2 \lambda_{c,0}^2}, \quad \left(R^{\pm}\right)^3 = A^{\pm}\xi_{a,0}^{\pm}\xi_{b,0}^{\pm}\xi_{c,0}^{\pm}. \tag{5}$$

However, due to inhomogeneities, a solid is homogeneous over a finite length $L$ only. In this case the actual correlation length $\xi_i^{\pm}$, cannot grow beyond $L$ as $t \to 0$, and the transition is rounded. Due to this finite size effect, the specific heat peak occurs at a temperature $T_p$ shifted from the homogeneous system by an amount $L^{-1/\nu}$. To illustrate the relevance of the inhomogeneity induced finite size effect we displayed in Fig.4 the temperature dependence of the specific heat coefficient $c/T$ of $YBa_2Cu_3O_{7-\delta}$. To set the scale we note that in $^4He$, falling into the same universality class, there are no strains and the only significant impurity is $^3He$, which occurs naturally only to the extent of 1 part in $10^7$. For this reasons the critical properties can be probed down to $|t| \approx 10^{-9}$. In Fig.4 we indicated the resulting critical behavior in terms of the dashed and solid lines. Thus there is an intermediate temperature region, where the data fall nearly on the respective straight lines, bounded by the regimes where the finite size effect dominates and the region where the contribution of the lattice degrees of freedom leads to deviations. A detailed fine size analysis points to rather microscopic inhomogeneities with a length scale $L$ ranging from 300 to 400 A [1]. For this reason, the evidence for 3D-XY critical point behavior is restricted to an intermediate temperature regime around $T_p$. Note that the universal relation (5) should apply for any $T_c(x,y)>0$, while the critical amplitudes depend on $x$ and $y$. Using $A^+ = 8.4\,10^{20}$ cm$^{-3}$ derived from the data shown in Fig.4, $\lambda_{a0} = 1153$ A, $\lambda_{b0} = 968$ A and $\lambda_{c0} = 8705$ A, derived from magnetic torque measurements on a sample with $T_c = 91.7$ K [26], as well as the universal numbers $A^+/A^- = 1.07$ and $R^- = 0.59$, we obtain $T_c = 88.2$ K. Thus, in nearly optimum doped $YBa_2Cu_3O_{7-\delta}$ the universal relation is remarkably well satisfied. However, due to the existence of 2D-QSI and 3D-QSN critical points and the associated dimensional crossover phenomena, it may be difficult to enter the regime where 3D-XY fluctuations dominate on the entire $T_c(x,y) > 0$ surface. Nevertheless, important implications emerge: Because on this surface thermal fluctuations dominate, pairing and pair condensation do not occur simultaneously. $T_c$ and the critical amplitudes of penetration depts. and specific heat are not independent but related by Eq.(5). Concerning the isotope effect it then follows that

$$\beta_{T_c} = \sum_{i=a,b,c} \beta_{1/\lambda_{i0}^2} - \beta_{A^-}, \quad (6)$$

where $\beta_B = -(M/T_c)\, dT_c/dM$ is the isotope coefficient of property B. Thus, the isotope effect does not provide information on the pairing mechanism. Moreover, the empirical fact that $\beta_{T_c}$ is very small close to optimum doping [27] implies that the combined effects on penetration depths and specific heat nearly cancel in this regime.

Next we turn to the line of quantum phase transitions $y_c(x)$ where $T_c$ vanishes (see Fig.3). Close to 2D-QSI criticality various properties are not independent but related by [1,2]

$$T_c = \frac{\Phi_0^2 R_2}{16\pi^3 k_B} \frac{d_s}{\lambda_{ab}^2(0)} \propto n_s^{\square}(0) \propto 1/\gamma_{T_c}^z \propto 1/\gamma_{T=0}^z \propto \delta^{\bar{\nu}z}. \quad (7)$$

$\lambda_{ab}(0)$ is the zero temperature in-plane penetration depth, $n_s^{\square}(0) \propto d_s/\lambda_{ab}^2(0)$ the aerial superfluid number density, $\gamma_{T_c}$ and $\gamma_{T=0}$ denote the anisotropy parameters evaluated at $T = T_c$ and $T = 0$, respectively, $z$ is the dynamic and $\bar{\nu}$ the correlation length critical exponent of the 2D-QSI transition. $\delta = x - x_u,\, y_c - y$ measures the distance from the 2D-QSI critical point and $R_2$ is a universal number. Since $T_c \propto n_s^{\square}(0)$ is a characteristic 2D property [1,2] it also applies to the onset of superfluidity in $^4$He films adsorbed on disordered substrates, where it is well confirmed [28]. A great deal of experimental work has also been done in cuprates on the so called Uemura plot, revealing an empirical correlation between $T_c$ and $1/\lambda_{ab}^2(0)$ [29]. As shown in Fig.5, approaching 2D-QSI criticality, the data of a given family tends to fall on straight line, consistent with a linear relationship between $T_c$ and $1/\lambda_{ab}^2(0)$. Thus, the difference in the slope of the dashed and solid line reflect the family dependent value of $d_s$, the thickness of the sheets, becoming independent in the 2D limit. The relevance of $d_s$ was also confirmed in terms of the relationship between the isotope effect on $T_c$ and $1/\lambda_{ab}^2(0)$ [27]. Here it enters according to Eq.(7) in terms of

$$\beta_{T_c} = \beta_{d_s} + \beta_{1/\lambda_{ab}^2(0)}, \tag{8}$$

valid close to 2D-QSI criticality. Clearly, in this regime the scaling relations (6) and (8) have to match.

We are now prepared to discuss the empirical correlations and phase diagrams from the point of view of critical phenomena. Close to quantum criticality $T_c$ vanishes as $T_c \propto \delta^{z\bar{\nu}}$, where $\delta$ measures the distance from the quantum critical point. The empirical relation (1) then implies 2D-QSI and 3D-QSN transitions transition with $z\bar{\nu}=1$, while empirical correlations for the anisotropy (Eqs.(2) and (3)), together with the scaling relation (7), yield $\bar{\nu}=1$ at the 2D-QSI critical point. Thus, the empirical correlations point to a 2D-QSI transition with

$$2D-QSI: \quad z\bar{\nu}=1, \quad \bar{\nu}=1. \tag{9}$$

These estimates coincide with the theoretical prediction for a 2D disordered bosonic system with long-range Coulomb interactions, where $z=1$ and $\bar{\nu}\approx 1$ [33-35]. Here the loss of superfluidity is due to the localization of the pairs, which is ultimately responsible for the transition. Returning then to the phase diagram shown in Fig.3, we observe remarkable consistency between the empirical correlations and the critical properties of a 2D-QSI transition in the underdoped region of the line $y_c(x)$ for a variety of cuprates. It clearly reveals that the doping and substitution tuned flow to the 2D-QSI critical points is associated with a depression of $T_c$ and an enhancement of $\gamma_T$. Consequently, whenever a 2D-QSI transition is approached, a no vanishing $T_c$ is inevitably associated with an anisotropic but 3D condensation mechanism, because $\gamma_T$ is finite (see Fig.2). This points unambiguously to the conclusion that theories formulated for a single $CuO_2$ plane cannot be the whole story.

Finally, we turn to the 3D-QSN transitions along the critical endline $y_c(x)$ (see Fig.3)). Here the scaling theory predicts [1,2]

$$T_c \propto \delta^{z\bar{\nu}} \propto \left(1/\lambda_{ab}^2(0)\right)^{\frac{z}{z+1}}, \quad \gamma|_{T=0} = c/T|_{T=0} \propto H^{(D-z)/2}, \tag{10}$$

where $\gamma$ denotes here the specific heat coefficient. Here the factor of proportionality in the relationship between $T_c$ and $1/\lambda_{ab}^2(0)$ is non universal. This behavior can be anticipated from the $\mu$SR data for $TlBa_2CuO_{6+\delta}$ [36], $Y_{0.2}Ca_{0.2}Ba_2(Cu_{1-y}Zn_y)_3O_{7-\delta}$ and $Tl_{0.5-y}Pb_{0.5+y}Sr_2Ca_{1-x}Y_xCu_2O_7$ [37]. Unfortunately, the data are too sparse to derive an estimate for the dynamic exponent $z$. A potential candidate for 3D-QSN criticality is the disordered d-wave superconductor to normal state transition at weak coupling considered by Herbut [38], with

$$\text{3D-QSN:} \quad z = 2, \quad \bar{\nu} = 1/2, \quad z\bar{\nu} = 1. \tag{11}$$

A property suited to shed light on the critical behavior of both, the 2D-QSI and 3D-QSN transition is the magnetic field dependence of the specific heat coefficient in the limit of zero temperature. From the scaling law (10) and the exponents listed in Eqs.(9) and (11) it is seen that for both transitions $\gamma|_{T=0} = c/T|_{T=0} \propto H^{1/2}$. Remarkably enough, this is what has been found in $La_{2-x}Sr_xCuO_4$ , irrespective of the doping level [39]. ]. Another quantity of interest is the condensation energy at $T = 0$. Close to 2D-QSI and 3D-QSN criticality it scales as

$$-E(T=0) \propto \delta^{\bar{\nu}(D+z)}. \tag{12}$$

Thus, along the hole concentration axis we obtain with the exponents listed in Eqs.(9) and (10),

$$-E(T=0) = a_{QSI}(p - p_u)^3 \propto T_c^3, \quad -E(T=0) = a_{QSN}(p_o - p)^{5/2} \propto T_c^{5/2}, \tag{13}$$

close to the 2D-QSI and 3D-QSN critical point, respectively. From Fig.6 it is seen that this behavior is remarkably consistent with the experimental data for $Y_{0.2}Ca_{0.2}Ba_2Cu_3O_{7-\delta}$ [40].

In conclusion, we have seen that the 2D-QSI transition has a rather wide and experimentally accessible critical region. For this reason we observed considerable and consistent evidence that it falls into the same universality class as the onset of superfluidity in $^4$He films in disordered media, corrected for the long-rangeness of the Coulomb interaction. The resulting critical exponents, are also consistent with the empirical relations and the observed magnetic field dependence of the specific heat coefficient in the limit of zero temperature and the doping dependence condensation energy. These properties also point to a 3D-QSN transition describing a d-wave superconductor to disordered metal transition at weak coupling. Here the disorder destroys superconductivity, while at the 2D-QSI transition it localizes the pairs and with that destroys superfluidity. Due to the existence of the 2D-QSI and 3D-QSN critical points, the detection of finite temperature 3D-XY critical behavior will be hampered, in addition to the finite size effects arising from inhomogeneities and the associated crossovers which reduce the temperature regime where thermal 3D-XY fluctuations dominate. In any case, our analysis clearly revealed that superconductivity in bulk cuprate superconductors is a genuine 3D phenomenon and that the interplay of anisotropy and superconductivity destroys the latter in the 2D limit. As a consequence, the universality of the empirical correlations reflect the flow to 2D-QSI and 3D-QSN criticality, tuned by doping, substitution etc. A detailed account of the flow from 2D-QSI to 3D-QSN criticality is a challenge for microscopic theories attempting to solve the puzzle of superconductivity in these materials. Although the mechanism for superconductivity in cuprates is not yet clear, essential constraints emerge from the existence of the quantum critical endpoints and lines. However, much experimental work remains to be done to fix the universality class of the 2D-QSI and particularly of the 3D-QSN critical points unambiguously.

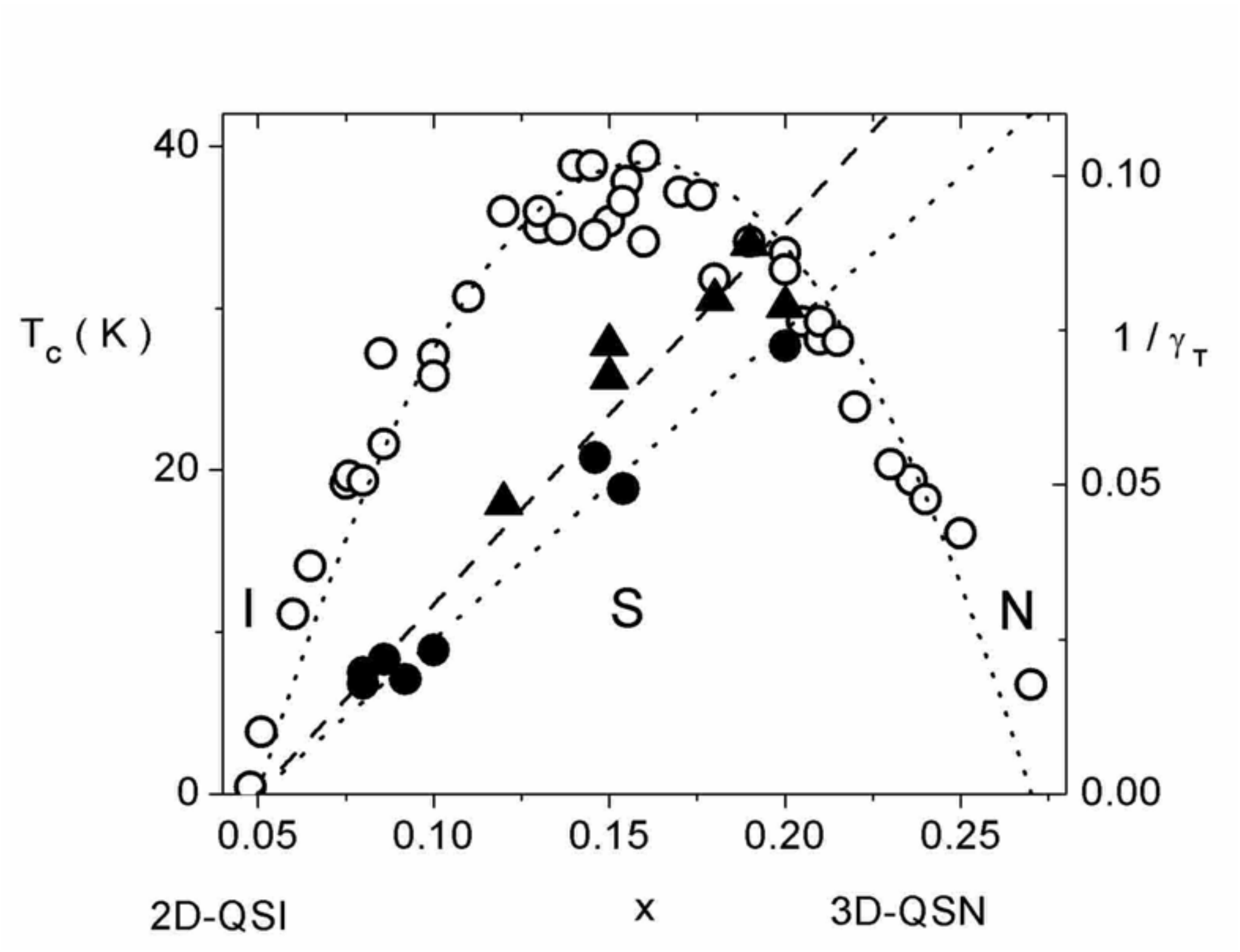

Fig.1:Variation of $T_c$ (○) [2-10] and $\gamma_T$ with $x$ for $La_{2-x}Sr_xCuO_4$. ● mark $\gamma_{T_c}$ [3,4,6,8] and ▲ $\gamma_{T=0}$ [10,11]. The solid curve is Eq.(1) with $T_{cm} = 39$ K. The dashed and dotted lines follow from Eq.(2) with $\gamma_{T=0,0} = 1.63$ and $\gamma_{T_c,0} = 2$ .

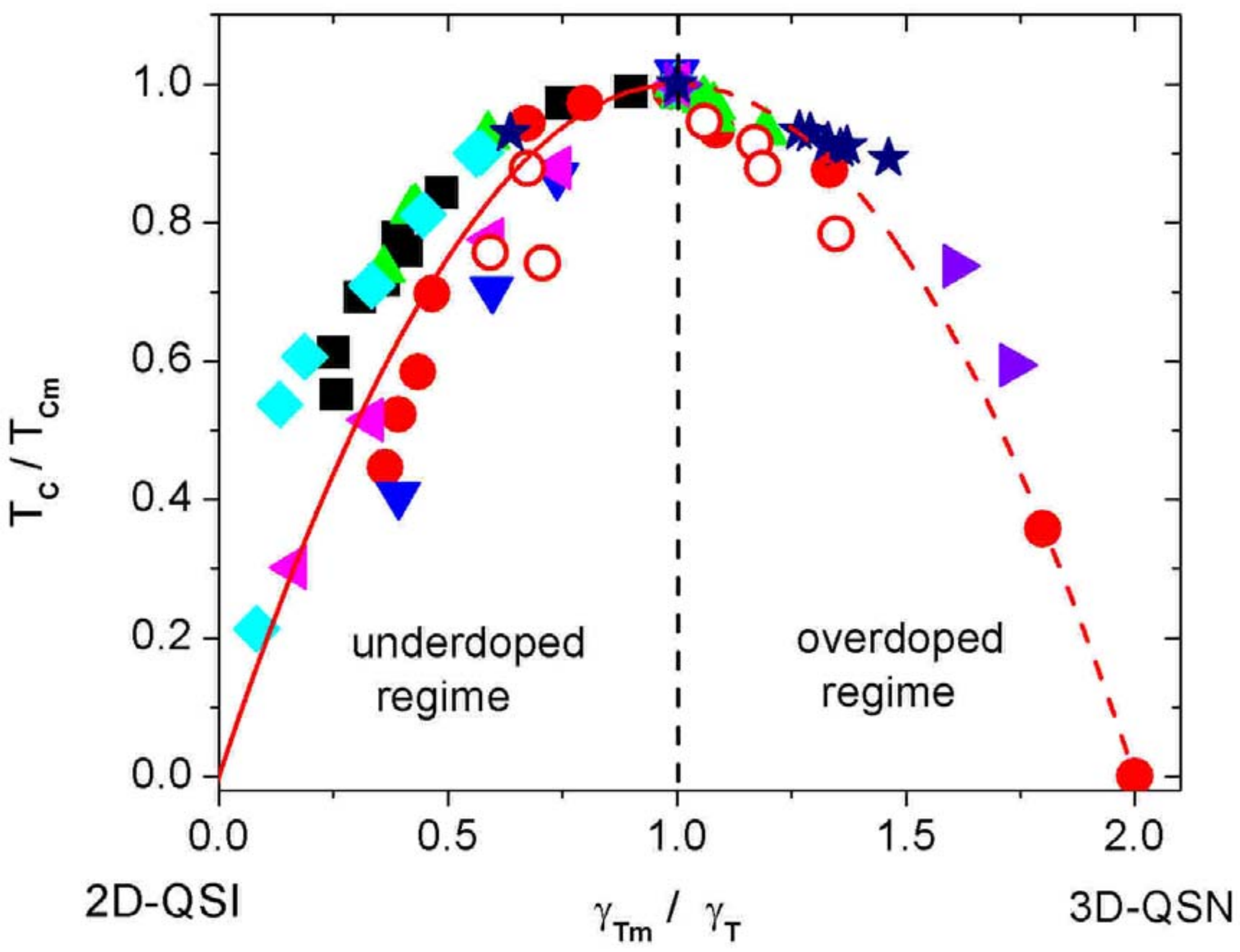

Fig.2: $T_c / T_{cm}$ versus $\gamma_{Tm} / \gamma_T$ for

$La_{2-x}Sr_xCuO_4$ (●, $T_{cm} = 37$ K, $\gamma_{T_{cm}} = 20$ [3,4,6,8], ○, $T_{cm} = 37K$ , $\gamma_{T=0m} = 14.9$ [10,11]),

$HgBa_2CuO_{4+\delta}$ (▲, $T_{cm} = 95.6$ K, $\gamma_{T_{cm}} = 27$ [17]),

$Bi_2Sr_2CaCu_2O_{8+\delta}$ ( ✶, $T_{cm} = 84.2$ K, $\gamma_{T_{cm}} = 133$ [18]),

$YBa_2Cu_3O_{7-\delta}$ (◆, $T_{cm} = 92.9$ K, $\gamma_{T_{cm}} = 8$ [19]),

$YBa_2(Cu_{1-y}Fe_y)_3O_{7-\delta}$ (■, $T_{cm} = 92.5$ K, $\gamma_{T_{cm}} = 9$ [20]),

$Y_{1-x}Pr_xBa_2Cu_3O_{7-\delta}$ (▼, $T_{cm} = 91$ K, $\gamma_{T_{cm}} = 9.3$ [21]),

$BiSr_2Ca_{1-y}Pr_yCu_2O_8$ (◀, $T_{cm} = 85.4$ K, $\gamma_{T=0m} = 94.3$, [22]) and

$YBa_2(Cu_{1-y}Zn_y)_3O_{7-\delta}$ (▶, $T_{cm} = 92.5$ K, $\gamma_{T=0m} = 9$ [23]).

The solid and dashed curves are Eq.(3), marking the flow from the maximum $T_{cm}$ to 2D-QSI and 3D-QSN criticality, respectively.

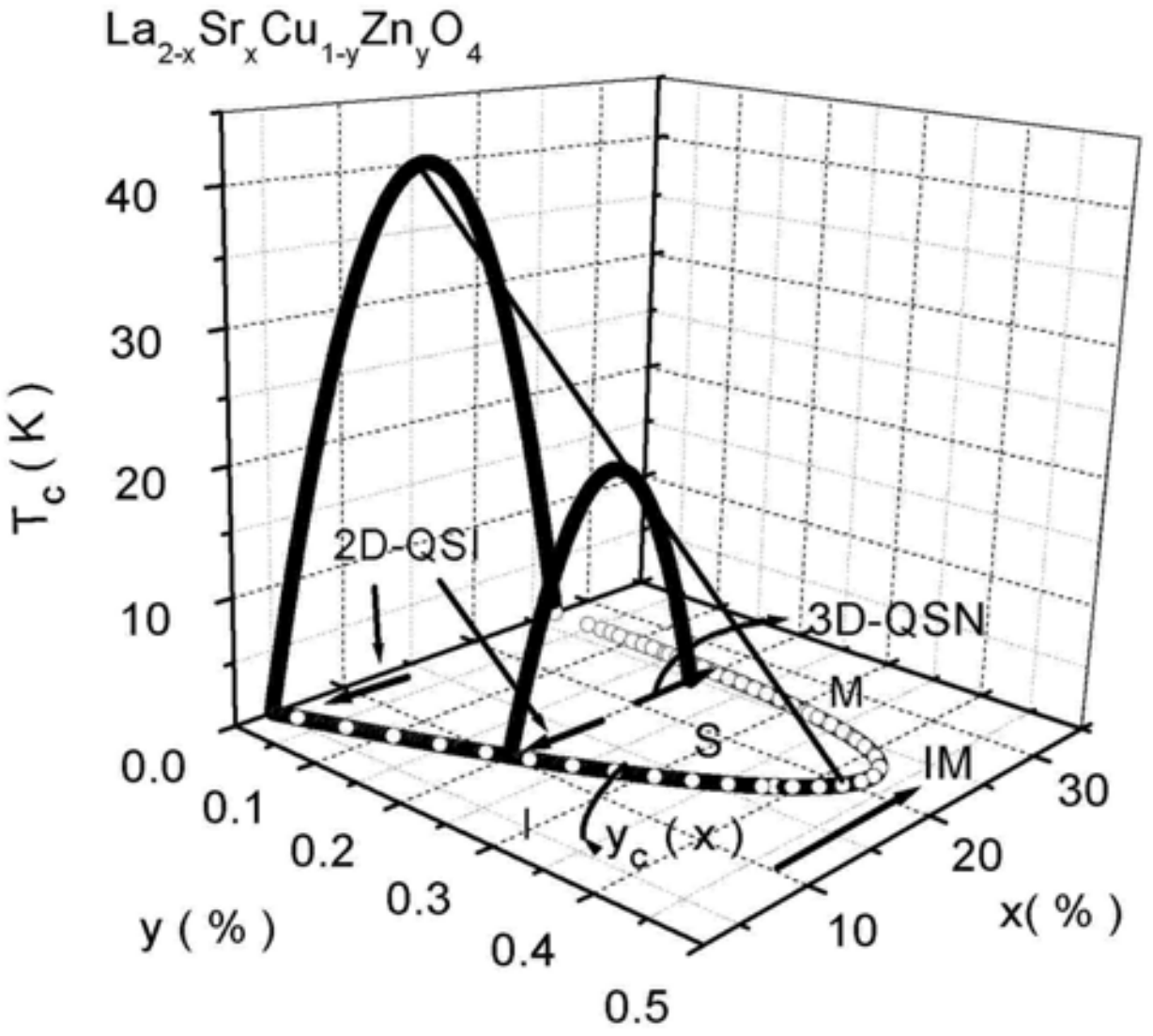

Fig.3: $T_c(x, y)$ phase diagram of $La_{2-x}Sr_xCu_{1-y}Zn_yO_4$ derived from the experimental data of Momono et al. [13]. $y_c(x)$ is a line of 2D-QSI and 3D-QSN transitions.

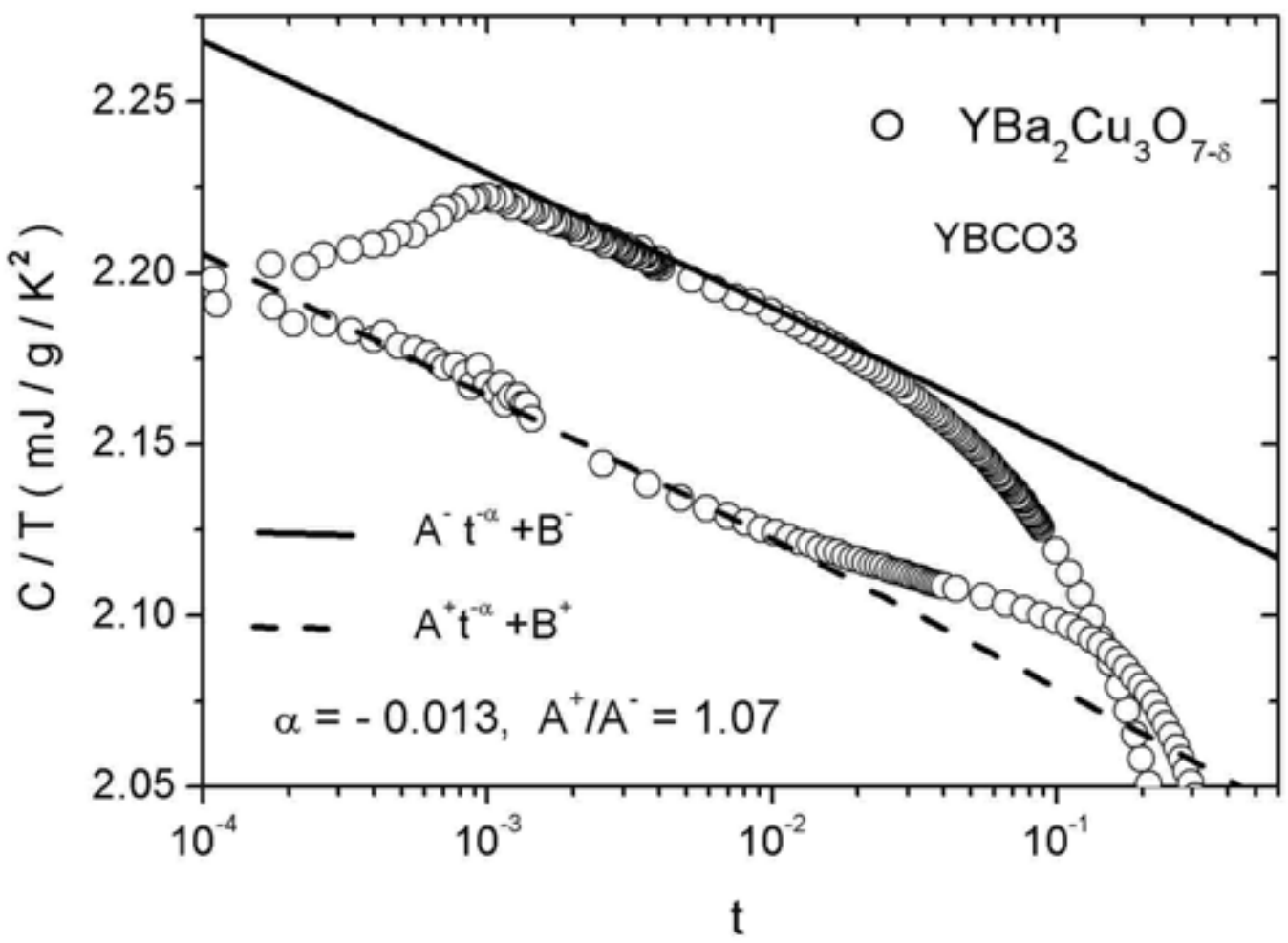

Fig.4: Specific heat coefficient $c/T$ versus $|t| = |1 - T/T_c|$ of $YBa_2Cu_3O_{7-\delta}$ with $T_c = 92.12$ K.. ○: Experimental data taken from [25].

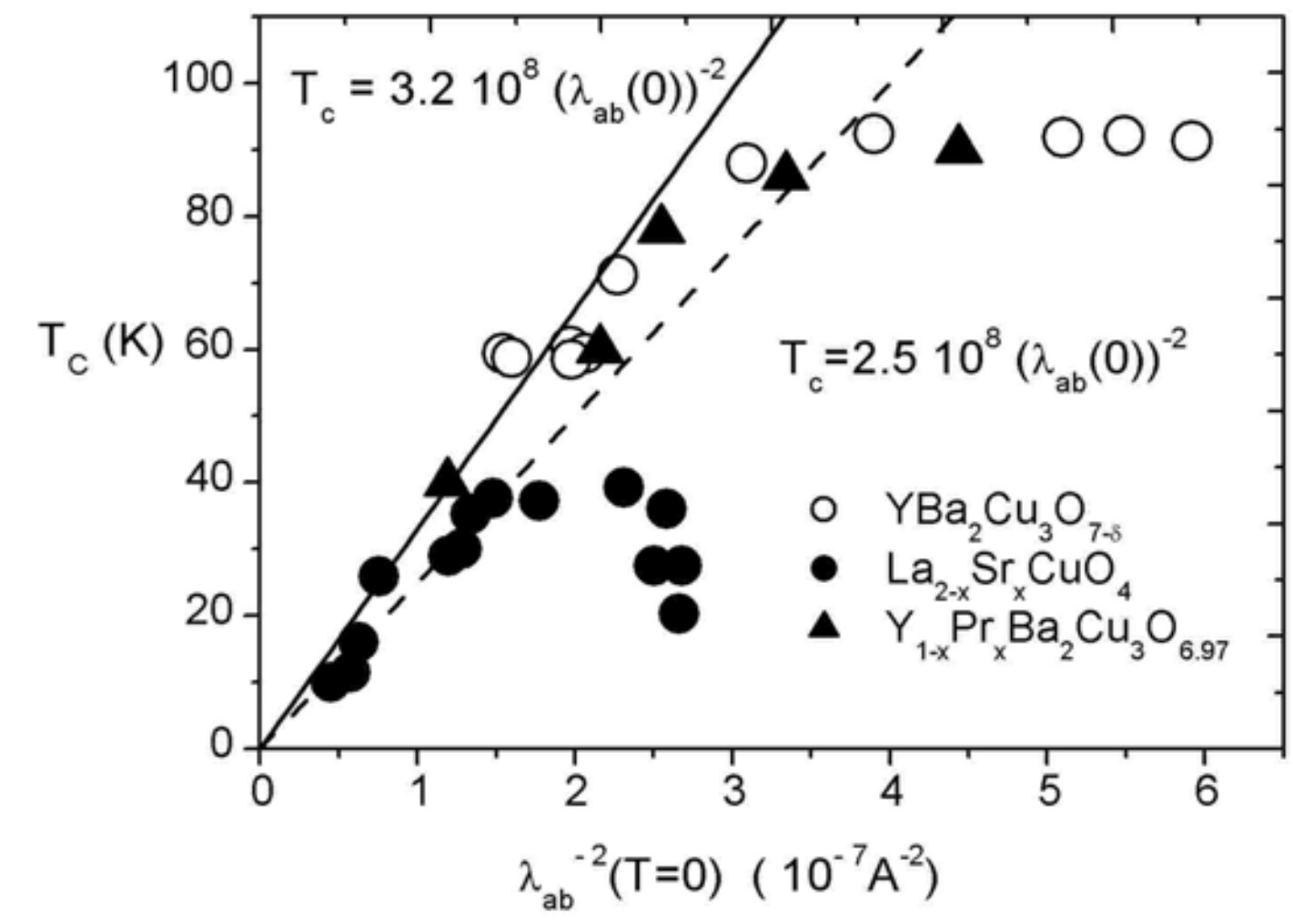

Fig.5: $T_c$ versus $\lambda_{ab}^{-2}\left(T=0\right)$ for $La_{2-x}Sr_xCuO_4$ [11,29,30], $YBa_2Cu_3O_{7-\delta}$ [31] and $Y_{1-x}Pr_xBa_2Cu_3O_{7-\delta}$ [32]. The solid and dashed lines are Eq.(7) and indicate the variation in $d_s$ .

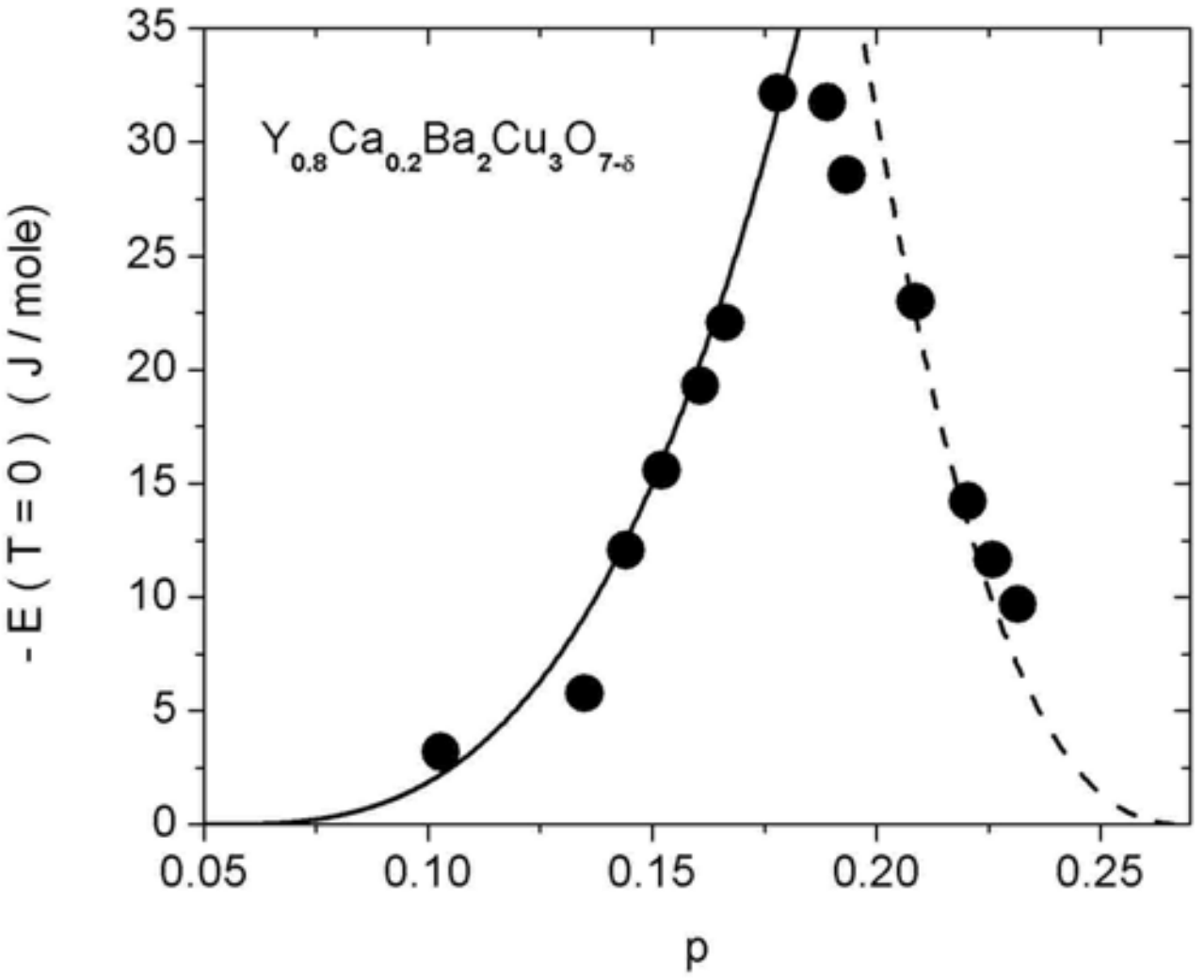

Fig.6: Condensation energy $E\left(T=0\right)$ versus hole concentration $p$ for $Y_{0.2}Ca_{0.2}Ba_2Cu_3O_{7-\delta}$. ● taken from [40]. The solid and dashed curves are Eqs.(13) with $a_{QSI}=15000$ and $a_{QSN}=24000$.